\documentstyle[prl,aps,floats]{revtex}
\input{epsf}
\newcommand{\be}{\begin{equation}}
\newcommand{\ee}{\end{equation}}
\newcommand{\bea}{\begin{eqnarray}}
\newcommand{\eea}{\end{eqnarray}}
\newcommand{\bean}{\begin{eqnarray*}}
\newcommand{\eean}{\end{eqnarray*}}
\font\upright=cmu10 scaled\magstep1
\font\sans=cmss10
\newcommand{\ssf}{\sans}
\newcommand{\stroke}{\vrule height8pt width0.4pt depth-0.1pt}
\newcommand{\Z}{\hbox{\upright\rlap{\ssf Z}\kern 2.7pt {\ssf Z}}}

\newcommand{\C}{{\rlap{\rlap{C}\kern 3.8pt\stroke}\phantom{C}}}
\newcommand{\R}{\hbox{\upright\rlap{I}\kern 1.7pt R}}
\newcommand{\CP}{\C{\upright\rlap{I}\kern 1.5pt P}}
\newcommand{\PP}{\hbox{\upright\rlap{I}\kern 1.5pt P}}

\begin{document} 
\twocolumn[\hsize\textwidth\columnwidth\hsize\csname@twocolumnfalse\endcsname
\begin{flushright}   
\ \vskip -0cm{\normalsize To appear in Phys. Rev. E.}\\
\end{flushright}
\title{On the Stability of Knots in Excitable Media}
\author{Paul M. Sutcliffe{$^{*}$} and Arthur T. Winfree{$^\dagger$}}
\address{
{$^*$} Institute of Mathematics, University of Kent,
Canterbury CT2 7NF, U.K.\\
{$^\dagger$} Department of Ecology and Evolutionary Biology,
University of Arizona, Tucson AZ 85721, USA.\\
}
\maketitle
\begin{abstract}
Through extensive numerical simulations we investigate the
evolution of knotted and linked vortices in the FitzHugh-Nagumo model.
On medium time scales, of the order of a hundred times the vortex
rotation period, knots simultaneously translate and precess with
very little change of shape. However, on long time scales
we find that knots evolve in a more complicated manner, with
particular arcs expanding and contracting, producing substantial
variations in the total length. The topology of a knot
is preserved during the evolution and after several thousand
vortex rotation periods the knot appears to approach an asymptotic
state. Furthermore, this asymptotic state is dependent upon the
initial conditions and suggests that, even within a given
topology, a host of meta-stable 
configurations exist, rather than a unique stable solution. 
We discuss a possible mechanism for the observed evolution,
associated with the impact of higher frequency wavefronts emanating from 
parts of the knot which are more twisted than the expanding arcs.\\
\end{abstract}

]

There are a wide variety of naturally occurring excitable media which
possess spiral wave vortices. Examples include chemical concentrations
in the Belousov-Zhabotinsky reaction 
\cite{Win1} and electrical depolarization waves
 in cardiac tissue \cite{Witetal}.
The vortices in this last example are of particular significance since
they are believed to play a vital role in ventricular fibrillation and
hence sudden cardiac death \cite{special}.
Both these systems, and many others, have a common mathematical description
in terms of nonlinear partial differential equations of 
reaction-diffusion type. In the case of cardiac tissue the simplest
continuous mathematical model is the FitzHugh-Nagumo equation, and it is
the investigation of dynamical three-dimensional solutions of this equation
which is the topic of this letter. 

More than twenty years ago it was conjectured that stable,
or at least persistent, three-dimensional
\hbox{solutions} (termed \lq organizing centres\rq) 
might exist in excitable media in which two-dimensional vortices
are embedded into three-dimensional space in such a way that they 
form knotted (or linked) vortex strings.
The anatomy of these objects was clarified in terms
of the topology of iso-concentration surfaces bordered by vortex strings
\cite{WS}.
 The hope was that the non-trivial
topology of a configuration,
perhaps aided by 
a short-range repulsive force
between vortex cores, or by an effect of phase 
twist along the vortex string,
 might provide a barrier to its decay \cite{WG}.
However, others \cite{Kri} argued against this optimism with the view
that curvature, tension and reconnection processes would ultimately lead
to the collapse and extinction of all knots.
A framework was proposed \cite{Kee,Bik} for thinking
about vortex string dynamics in the limiting case of slight
curvature and twist, but attempts to verify it 
were successful only in the strict limit of no twist \cite{Win2,Hen}.
Ultimately, to address
the fundamental issue of the existence of stable knots one must turn
to numerical \hbox{methods.}
 About a decade ago a number of preliminary numerical
investigations were performed \cite{Win2,Hen} 
which suggested that certain knotted
(and linked) configurations were stable, having a soliton-like behaviour in
which the knot moved through the medium as a rotating rigid body with
a constant shape.  However, due to computational constraints, such 
simulations were limited to time scales which never exceeded about 
one hundred
vortex rotation periods (often substantially less) and used 
very symmetric initial conditions.

In this letter we present the results of extensive numerical simulations
of a duration well beyond a thousand times the vortex period, and using
perturbed asymmetric initial conditions. We investigate several
knots and links and conclude that all appear to be meta-stable in the
sense that small perturbations produce dramatic changes in the evolution
over time scales of the order of thousands of vortex rotation periods.
The evolution is quite exotic, and very far from the simple curvature
 and tension-induced collapse suggested
previously. In all cases the topology of the knot (or link) is 
preserved during the evolution, as we observe no reconnection events.
Rather than a simple uniform {contraction} of the knot, which might be expected
as a result of tension, we find that a particular arc
of the knot both {expands} and contracts. 
After substantial variations in its total length the knot
eventually approaches a steady state. However, this state does not appear
to be unique and suggests that, even within a given
topology, a host of meta-stable configurations exist.
We discuss a possible mechanism for the observed evolution,
associated with the impact of higher frequency wavefronts emanating from 
parts of the knot which are more twisted than the expanding arcs.

The FitzHugh-Nagumo equations are given by
\be
\frac{\partial u}{\partial t}=\frac{(u-u^3/3-v)}{\epsilon}+\nabla^2 u,
\quad
\frac{\partial v}{\partial t}=\epsilon(u+\beta-\gamma v)
\label{fhn}
\ee
where $u(t,{\bf x})$ and $v(t,{\bf x})$ are both real fields
with $u$ the electric potential and $v$  the recovery 
variable associated with membrane channel conductivity.
We take the constants appearing in equation (\ref{fhn}) to have
the values $\epsilon=0.3,\beta=0.7,\gamma=0.5.$
 This choice 
of constants is non-generic and is motivated
by our aim of trying to find stable knots. This
set of values has a number of special properties which might be 
conducive to knot stability, such as the lack of meander of a 
two-dimensional vortex and the stability of an untwisted vortex ring
in three dimensions. See \cite{Win5} for a description of the
properties of a two-dimensional vortex as a function of the parameters
$\epsilon,\beta,\gamma.$

In two space dimensions the FitzHugh-Nagumo equations with these
parameter values has plane wave solutions which travel at a speed
$c=1.9$ and rotating vortex solutions
(often called spiral waves) with a period 
$T_0=11.2.$ The vortex solution has $u$ and $v$ wavefronts in the
form of an involute spiral with a wavelength $\lambda_0=cT_0=21.3.$
Geometrically this means that all lines which are perpendicular to the level
 curves of the field $u$ are tangent to a small circle of diameter 
$\lambda_0/\pi.$ This circle represents the vortex core and is the region
in space in which the gradients of the $u$ and $v$ fields differ substantially
from being parallel. For later use it is convenient to define the quantity
\be
\Phi=|{\bf \nabla} u\times{\bf \nabla} v|
\label{core}
\ee
which is highly localized at the vortex core.

We solve equations (\ref{fhn}) in three-space dimensions using an
explicit finite difference scheme which is accurate to second order
in the spatial derivatives and first order in the time derivative.
Although this scheme appears very simplistic, it appears that the
nature of these equations is such that more sophisticated or higher
order algorithms do not lead to substantial gains in efficiency or
accuracy, although this can be achieved if one is willing to
modify the FitzHugh-Nagumo equations to a form
designed specifically for the applicability of a more efficient 
numerical approach \cite{Bar}. 
For our simulations we use a grid containing
$201^3$ points and a lattice spacing $\delta x=0.5,$ so that our
spatial coordinates are confined to the range $-50\le x_i \le 50.$
The time step used is $\delta t=0.02.$ In the
$x_1$ and $x_2$ directions we apply Neumann boundary conditions and
in the $x_3$ direction the boundary conditions are periodic. The
selection of the $x_3$ direction as periodic is because we shall
orient our knots so that they initially translate as rigid bodies 
moving in the $x_3$ direction, and we do not wish to impede their motion.

We create initial conditions which form knotted vortex strings by 
making use of complex curves as described in \cite{PW} and is similar
to the approach used in \cite{BS1}
 for the study of knotted topological solitons.
Recall that a knot may be written as the intersection of
a complex curve ${\cal C}$ with the unit 3-sphere $S^3.$ 
Here $S^3$ should be thought of as compactified three-dimensional 
Euclidean space, with the explicit
coordinates given by stereographic projection 
\be
Z_0=\frac{2(x_1+ix_2)}{1+r^2},\quad
Z_1=\frac{r^2-1+2ix_3}{1+r^2}
\label{stereo}
\ee
where $r$ is the Euclidean distance from the origin and $Z_0$ and
$Z_1$ are two complex coordinates satisfying $|Z_0|^2+|Z_1|^2=1$
and hence parameterize $S^3.$ With this identification the knot
 is the one-dimensional locus in space of the complex
curve ${\cal C}$ with coordinates $Z_0$ and $Z_1.$ As an example,
to represent the $(m:n)$ torus knot we take 
${\cal C}=Z_1^m-Z_0^n,$ where for later convenience we have
identified ${\cal C}$ with its zero set.
If ${\cal C}$ has $p$ factors then it describes an object with  $p$
components and hence this formalism can also be used to describe 
 disconnected knots as well as links. 

For a given knot (or link) we create initial conditions for the
 fields $u$ and $v$ from the associated curve ${\cal C}$
through the prescription
\be
u=\Lambda_1 \mbox{Real}({\cal C})+u_*,\quad
v=\Lambda_2 \mbox{Imag}({\cal C})+v_*.
\ee
Here $\Lambda_1$ and $\Lambda_2$ are two real constants 
(taken to be $\Lambda_1=2,\Lambda_2=1$)
which are
used to scale the initial conditions so that they cover the range
of the excitation-recovery loop in $(u,v)$ space associated with
the ODE part of equation (\ref{fhn}). 
The constants $u_*=v_*=-0.4$
are the values which can be roughly attributed to the vortex core.
\begin{figure}
\begin{center}
\epsfxsize=9cm\epsfysize=9cm\epsffile{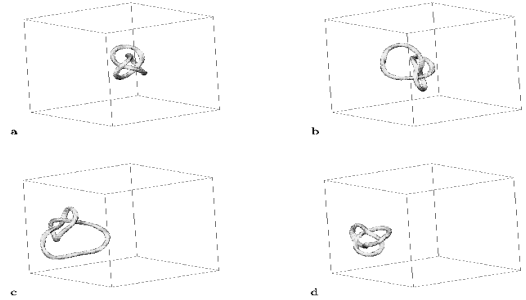}
\ \vskip -2cm
\caption{The core of the trefoil knot at times $t=500,5000,10000,40000.$ }
\label{tref_expand}
\end{center}
\end{figure}
\begin{figure}
\begin{center}
\epsfxsize=9cm\epsfysize=6cm\epsffile{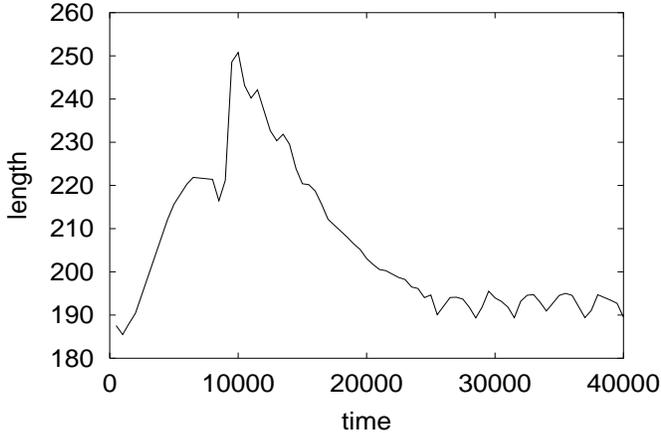}
\caption{
The length of the trefoil knot as a function of time.
}
\end{center}
\end{figure}

The simplest non-trivial knot is the trefoil knot, given by the
curve ${\cal C}=Z_1^2-Z_0^3.$ We use this in the above prescription
to obtain our initial conditions.
In fig.~\ref{tref_expand} we plot the isosurface $\Phi=0.01$,
which indicates the core of the vortex string, at the times
$t=500,5000,10000,40000.$ Note that 
at time $t=0$ this isosurface vanishes,
 since the initial conditions do not produce the
vortex string itself but only seed the field configuration which
will form into a vortex string after a time scale of the order of
ten vortex periods. In fig.~1a the symmetric trefoil knot has clearly
formed. In fact the knot forms at a much earlier time, but it is slightly
larger and quickly shrinks to this size. The knot moves in the
$x_3$ direction (towards the back left-hand-side of the box
in the figure), with little
change of shape, and at a speed of approximately
$c/80,$ where $c$ denotes the wavefront speed given earlier. 
The knot also 
rotates around the $x_3$-axis with a period of around $160T_0,$ where
$T_0$ is the vortex period given above.
Fig.~1b reveals that the knot has drifted slightly
away from the $x_3$-axis and one of the three lobes has expanded 
in comparison to the other two. Although the initial configuration has
a cyclic $C_3$ symmetry, the cubic grid, and more importantly its boundary,
breaks this symmetry and allows an asymmetric instability to develop.
The larger lobe continues to expand, fig.~1c, and now the knot no longer simply
translates in the $x_3$-direction, but rotates and follows a complicated
path in space. Though still preserving the topology
of a trefoil, the knot is now better viewed as a large expanding ring 
with a small knot tied in it.
Eventually the expansion of the arc stops and a contraction begins.
By fig.~1d the knot has regained its more symmetric form and has a similar
length as before the expansion, but now it appears to be an 
asymptotic state. This can be seen by computing the length as a function
of time, which is displayed in fig.~2. 

To understand a possible mechanism responsible for the expansion of one arc of
the knot and the subsequent contraction to a steady state
we need to recall two facts. First, analytical and numerical
work shows that a straight and uniformly twisted vortex line has a
period which is slightly less than that of the two-dimensional vortex,
or equivalently the untwisted vortex line \cite{MPR,PAK}. 
Here twist refers to the variation of the phase in the complex $(u,v)$-plane
as one moves along the vortex string. Second, it is known that
for a system of two vortices in which the vortices have different periods
(for example, as arises in a model with spatially varying parameters)
then the collision interface, which is the point at which the spiral wavefronts
from the two vortices meet and annihilate, gradually moves towards the
vortex with the larger period. 
In the absence of dispersion the collision interface moves at a speed
$\widehat c=c|T_1-T_2|/(T_1+T_2)$ where $T_1$ and $T_2$ are the periods
of the two vortices.
Eventually the collision interface reaches
the core of the larger period vortex and it gets slapped away by the
higher frequency wavefronts emanating from the shorter period vortex
\cite{YKPP,Vin}.   
Combining these two facts we see that a reasonable explanation for the
expanding arc is that the more knotted part has a greater local twist rate
than at least some part of the large expanding ring, so its period is slightly
less and this results in its higher frequency wavefronts slapping away
the ring and producing its expansion. 
This slapping mechanism is discussed in \cite{Win3} in the context
of stabilizing a knot against contraction.
To check this hypothesis we have
examined the collision interface by 
taking a slice through the configuration in which the expanding arc and
most parts of the knot pass almost perpendicularly through the selected
plane. This reveals that the wavefront produced by the
tightly knotted cores impacts almost on top of the core of the expanding
arc, in agreement with our hypothesis for slapping induced
expansion. The details will be presented elsewhere \cite{SW2}.

The simplest example, in which the above issues regarding stability can be
investigated, is to study two rings linked once. Consider the 
complex curve ${\cal C}=Z_1^2-Z_0^2-\mu Z_0,$ where
$\mu$ is a real parameter. If $\mu=0$ then this curve is associated with
two identical rings which each contain one full twist and are linked once.
This configuration has a $C_2$ symmetry corresponding to a rotation
by $180^\circ$ around the $x_3$-axis. The link formed from this initial
 condition moves along the $x_3$-axis as a rotating rigid structure
and shows no sign of instability even upto $t=20000.$ The reason this example
differs from the trefoil knot in this respect is that a $C_2$ symmetry
is clearly more compatible with the cubic lattice (and boundary) 
of the numerical grid than the $C_3$ symmetry of the trefoil knot.
For this example we therefore require an explicit perturbation to test
the stability of this link. This is achieved by setting $\mu$ to be non-zero
in the above curve, which distorts one of the rings,  making it larger
than the other and hence breaking the $C_2$ symmetry. The results of
a numerical evolution with $\mu=0.5$ are displayed in fig.~3, where
we plot the vortex cores ($\Phi=0.01$ isosurface) at times 
$t=200, 6200, 10200, 15200.$
The larger of the two rings initially contracts but this is followed
by an expansion which yields an asymptotic state in which the larger
ring has a length similar to that in the perturbed initial condition.
In fig.~4 we plot the lengths of the small ring (bottom curve) and
the large ring (top curve) as a function of time.
From this figure it can be seen that the length of the small ring
remains almost constant and an asymptotic state has been reached
which is certainly very different from the unperturbed solution ($\mu=0$)
in which both rings have an equal length. 
 By examination of the collision
interface we again verify that the wavefronts from the small ring
impact on the vortex core of the large ring.
Moreover, an examination of the twist along each of the rings reveals
that the small ring has a roughly constant positive twist along its
length, but the large ring has a substantial variation in its twist rate,
containing regions of {\sl negative} twist even though the total twist
along its length sums to one full turn in the positive direction.
The fact that such a highly non-trivial distribution of twist
occurs in an apparently asymptotic state is further evidence that
a variety of meta-stable configurations exist in which the relative spatial
distribution of the strings is in equilibrium under the action of several
complicated forces in which the rate of twisting plays a vital role. 
\begin{figure}
\begin{center}
\epsfxsize=9cm\epsfysize=9cm\epsffile{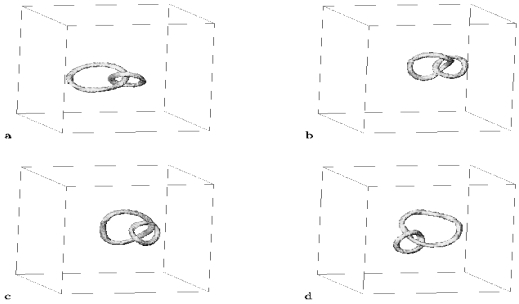}
\ \vskip -2cm
\caption{The core of the perturbed 
linked rings at times $t=200, 6200, 10200, 15200.$}
\label{2link1def}
\end{center}
\end{figure}
\begin{figure}
\begin{center}
\epsfxsize=9cm\epsfysize=6cm\epsffile{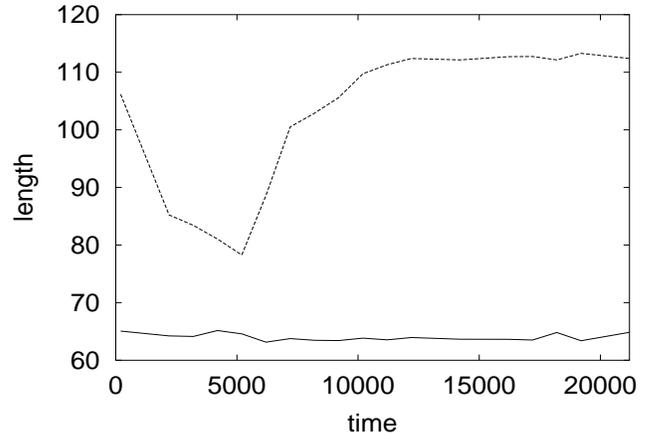}
\caption{
The length of the small ring (bottom curve) and the large ring
(top curve) as a function of time for the perturbed linked rings.
}
\label{2link1def2}
\end{center}
\end{figure}
To summarize, we have found a novel dynamical behaviour of knotted vortex
strings in the FitzHugh-Nagumo model with parameter values chosen
to minimize any knot instabilities. 
It would be interesting to determine if 
our results are generic for the FitzHugh-Nagumo model with other
parameter values and also for other excitable media. In fact, there
is already some evidence for this in the initial expansion of a trefoil knot
in a medium with equal diffusion of both reactants \cite{Win4}, but
this example was regarded as an unexplained peculiarity at the time
and simulations could not be performed for the length of time required
to observe the full expansion and approach to an asymptotic state that
we have described in this letter.
It would certainly be worthwhile performing extensive 
numerical investigations, over very long time scales,
 on a variety of equations modelling
different excitable media.

The construction of knotted vortex strings in laboratory experiments
on excitable media would be of significant interest, though it is
unlikely that the full evolution described in this letter could be studied
in this setting since the typical lifetime of vortices in current
experiments is limited to less than a hundred vortex periods.

Finally, the interaction and 
scattering of two initially well separated knots is also worthy
of investigation.

\end{document}